\documentstyle[preprint,aps]{revtex}
\begin{document}
\draft

\title{Prediction of an undimerized, insulating, antiferromagnetic
ground-state in halogen-bridged linear-chain Ni compounds}
\author{V. I. Anisimov, R. C. Albers, and J. M. Wills}
\address{Los Alamos National Laboratory, Los Alamos, New Mexico 87545}
\author{M. Alouani and J. W. Wilkins}
\address{Department of Physics, The Ohio State University,
Columbus OH 43210-1368}
\date{October 1994}
\maketitle

\begin{abstract}
A parameter-free, mean-field, multi-orbital Hubbard model with
nonspherical Coulomb and exchange interactions, implemented around
all-electron local-density approximation (LDA) calculations, correctly
predicts the band-gap energy, the absence of dimerization, and the
antiferromagnetic ground state of halogen-bridged linear-chain Ni
compounds.  This approach also reproduces the insulating ground state
and dimerization in PtX linear-chain compounds in agreement with
experiment and previous calculations.
\end{abstract}


\narrowtext
The halogen-bridged transition-metal linear-chain compounds, referred
to as MX compounds because of their alternating transition-metal atoms
M and halogen atoms X, form weakly coupled linear-chain-like
structures, and are of considerable interest
\cite{keller82,okamoto90,okamoto91,okamoto92,dono91,scott92,yamashita93,alouani92,alouani93,huang,bishop,gammel,kpti,larsen77,nicl}
due to a rich and accessible phase diagram
\cite{okamoto90,okamoto91,okamoto92,dono91,scott92,yamashita93,alouani92,alouani93,huang,bishop,gammel}.
More specifically, experimentalists can now tune the strength of the
charge-density wave (CDW) and spin-density wave (SDW)
\cite{okamoto92,yamashita93} and induce phase transitions from an CDW to
an SDW state \cite{okamoto91}.  First-principles density-functional
calculations of neutral PtX compounds have shown clearly that (1) the
mechanism of the dimerization and the insulating ground state in the
MX chain systems is due to an electron-phonon coupling along the
chain, and (2) the ligand structure is essential for the correct
description of ground-state properties \cite{alouani92}.  New
high-quality data on single crystals are challenging our understanding
of the electronic structure of these low-dimensional systems
\cite{okamoto90,okamoto91,okamoto92,dono91,scott92,yamashita93}.
The Ni MX compounds, which are normally in an SDW
state, change to a mixed SDW/CDW state by a simple substitution of the
counterion Cl$^-$ with perchlorite ClO$_4^-$ \cite{yamashita93}.  
In a recent Peierls-Hubbard model study, local defect
excitations in NiBr are found also to exhibit lattice distortion relative 
to the undistorted ground state \cite{huang}.
It is intersting to mention that 
in the Ni compounds the electronic structure at the vicinity of Fermi level 
is the one dimensional equivalent of the undoped cuprate superconductor 
materials, where
the local spin-polarized density approximation (LSDA) calculations predicted 
a non-magnetic metallic state instead of the observed 
antiferromagnetic insulating state. Here again the LSDA failed to reproduce
the ground state properties of the Ni compounds.

In this Letter we study the electronic properties of both CDW and SDW
chains by means of a newly developed mean-field, multi-orbital Hubbard
model, called ``LDA + U'', that has nonspherical Coulomb and exchange
interactions and is implemented around all-electron local-density
approximation (LDA) calculations with a linear muffin-tin orbital
basis-set \cite{anisimov93}.  This model enables us to extend our
previously successful LDA study of CDW MX systems
\cite{alouani92,alouani93} to also, for the first time, explain from
first-principles the magnetic SDW systems.  It predicts that the
ground state for the NiX (X= Cl or Br) compounds is undimerized,
insulating, and a low-spin antiferromagnetic, in good agreement with
experiment \cite{yamashita93}.  The large octahedral ligand-field
splitting $\Delta_0 = e_g -t_{2g} \approx 4$ eV produces a low-spin
electronic configuration $(t_{2g})^6 (e_g)^1$ of the Ni$^{+3}$.  It
also predicts that the ground state of the PtX (X=Cl or I) linear-chain 
compounds
is insulating and dimerized in agreement with our earlier calculations
\cite{alouani92}.  We have to stress that our calculations are based
on a parameter-free model that uses a constrained LDA determination
for the Coulomb and exchange interactions.

All the results presented in this Letter
have been obtained using the ``LDA + U'' method, which is extensively
described elsewhere \cite{anisimov93}.  Here we stress that the main
feature of the model is to introduce a discontinuity in the LDA
one-electron potential for integer change in orbital occupancy.  The
nonsphericity of the Coulomb and exchange interactions, i.e., the
$d$-orbital occupation dependence on the orbital angular-momentum
quantum number $m$ and $m^\prime$, is important for a good description
of the magnetic moment of the MX chains.

The $d_{3z^2-r^2}$ orbital, whose polarization dominates the magnetic
moment, is also important in controlling the bonding, the
dimerization, and the magnetism in these systems.  The total energy
and potentials include the exchange and the nonsphericity of the
Coulomb $d$-$d$ interaction:
\begin{eqnarray}
 E  = E_{LDA}-[UN(N-1)/2-JN(N-2)/4] 
 + {1\over 2}\sum_{m,m^\prime,\sigma}U_{mm^\prime} n_{m\sigma}
n_{m^\prime -\sigma} \nonumber \\ 
 + {1 \over 2}\sum_{m \neq m^\prime , m^\prime ,\sigma}
(U_{mm^\prime}-J_{mm^\prime})n_{m\sigma}n_{m^\prime \sigma} 
\end{eqnarray}

The screened Coulomb $U$ and exchange $J$ parameters are calculated
self-consistently in the supercell approximation as described in
\cite{ag}.  The orbital-dependent one-electron potential is
given by the derivative of Eq. (1) with respect to the orbital occupancy
$n_{m\sigma}$:
\begin{eqnarray}
V_{m\sigma}({\bf r})  = V_{LDA}({\bf r})+ 
 \sum_{m^\prime}(U_{mm^\prime}-U_{eff})n_{m^\prime -\sigma} \nonumber \\  \mbox{}
 +\sum_{m^\prime \neq m}(U_{mm^\prime}-J_{mm^\prime} -U_{eff})n_{m\sigma} 
+U_{eff}({1\over 2}-n_{m\sigma})-{1\over 4}J \; 
\end{eqnarray}
We define $U_{eff}=U-J/2$.  The matrices $U_{mm^\prime}$ and
$J_{mm^\prime}$ are calculated from the screened Slater integrals
$F^k$ ($F^0$, $F^2$, $F^4$ for d-electrons), which are determined
using the calculated values of $U$ and $J$ and the ratio of unscreened
$F^4 / F^2$ \cite{comm}.

Model calculations for the MX chain
compounds usually use one-dimensional one- or two-band systems in
which only the M $d_{3z^2-r^2}$ and the X $p_{z}$ orbitals are taken
into account \cite{gammel}.  Experimentally, MX compounds have
complicated structures of complex organic molecules ligands that are
perpendicular to the chain axis and bonded to the M atoms.  The
bonding orbitals of the ligands are strongly hybridized with the $d$
orbitals of the M atoms.  The MX compounds are divided into three
classes: (1) anionic, (2) cationic, and (3) neutral chains.  The
anionic chains have positive counterions so that the chains are
negatively charged; in contrast, the cationic chains have negative
counterions and are positively charged.  The neutral chains have no
counterions.  We have studied a case of an anionic CDW chain
K$_4$PtI$_4$PtI$_6$, and three cationic CDW and SDW chains
[M(chxn)$_2$][M(chxn)$_2$X$_2$]X$_4$ (chxn is 1,2-diaminocyclohexane,
C$_6$H$_{14}$N$_2$) with M= Ni (X= Br or Cl) or Pt with X= Cl.

{\it Cationic SDW chains} {[Ni(chxn)$_2$][Ni(chxn)$_2$X$_2$]X$_4$ }.
Depending on whether M is Pt or Ni, the associated MX chain shows a CDW
or a SDW character, respectively
\cite{okamoto90,okamoto91,okamoto92,dono91,scott92,yamashita93,alouani92,alouani93,bishop,gammel,kpti,larsen77,nicl}. 
For example, when MX is PtCl or PtBr, a Peierls instability is
observed and the system is non-magnetic. 
When M is Ni and the counterion is Cl or Br, these systems are
antiferromagnetic and exhibit no Peierls instability.  
\cite{okamoto90,yamashita93,nicl}.  We approximate the real structure
by a simpler model structure in order to make a direct
electronic-structure calculation feasible.  The simplest model
structure is to replace the (chxn)$_2$ by four ammonia units per
each M atom (see  Figure 1).
For the model crystal structure we compress the lattice along the $a$
axis so that the distance between ammonia on neighboring chains are
similar to those for ammonia-ligand compounds \cite{keller82}.
We emphasize that in our model all coordinates of the atoms are the
same as in real crystal structure except for the absence of the
C$_6$H$_{10}$ ring.  

The local spin-density approximation
(LSDA) alone can not reproduce the insulating magnetic ground-state
of the SDW  NiCl and NiBr systems.
The electronic structure in the vicinity of Fermi level is very
similar to that of the undoped cuprate superconductor materials, where
LSDA calculations also predict a non-magnetic metallic state instead
of the experimentally observed antiferromagnetic insulating state.
This effect results from the LSDA magnetic transition being driven by
the spin-polarization of a Stoner intraatomic exchange interaction $I$
(about 1 eV), instead of the much stronger Hubbard
interaction $U$ (about 8 eV).

Figure 2 shows the total density of states (DOS) and the energy dispersion 
along the chain
direction $\Gamma Z$ in the vicinity of Fermi level for the $U=J=0$ eV
case. Despite the presence of the half-filled Ni
$d_{3z^2-r^2}$ band at the Fermi level the Peierls mechanism is not
effective in opening the energy band gap. Instead, the small Ni-Ni
distance sufficiently enhances the anharmonic elastic potential
between Ni and Cl that the electron-phonon coupling (Peierls
mechanism) is rendered ineffective. \cite{alouani93}.

Figure 3 presents the Ni partial DOS at the vicinity of Fermi level
decomposed into the various symmetries of the $d$ channels for 
the calculated $U$ and $J$ as presented in Table 1 together with  
the dimerization, the
magnetization and the energy band gap for all the systems studied here.
The LDA+U  calculated  ground state remains undimerized 
and the  the Hubbard
term in Eq. (2) opens an energy band gap of 1.8 eV, in good agreement
with the experimental value of 1.9 eV \cite{nicl}.  Further, 
a low-spin antiferromagnetic state with a magnetic
moment of 0.66 $\mu_B$ is produced, which could be reduced by spin
fluctuations neglected in the present study. The experimental value of the
magnetic moment is not available at the present time.  
The low-spin antiferromagnetic state is due to the
large octahedral ligand-field splitting $\Delta_0 = e_g -t_{2g}
\approx 4$ eV which produces a low-spin $(t_{2g})^6 (e_g)^1$
electronic configuration of the Ni$^{+3}$ ion. 
Figure 3 shows that, with respect to the the DOS projected on the Ni site,
most of the weight of the
$(e_g)^1$ configuration comes from the $d_{3z^2-r^2}$ orbital.  
Accordingly most of the  contribution to the
spin magnetic moment is from the polarized $d_{3z^2-r^2}$ orbitals
located in the vicinity of the band gap, and that only 0.05 $\mu_B$ is
due to the polarization of the $d_{x^2-y^2}$ orbitals. In addition, the 
optical band gap of 1.8 eV between the 
occupied Ni $d_{3z^2-r^2}$ and  the unoccupied Pt $d_{x^2-y^2}$
is smaller than the antiferromagnetic gap of 2.3 eV arising from the
splitting of the $d_{3z^2-r^2}$ Pt bands.  A slightly smaller value of
$U$ makes these two gaps comparable.

The properties of the NiBr system are very similar to that of NiCl: an
antiferromagnetic insulator with a smaller gap of 1.28 eV
\cite{okamoto90,nicl}.  Our calculation produces a Coulomb parameter
$U$ = 6.9 eV, and an energy band gap of 1.6 eV with a spin magnetic
moment value of 0.63 $\mu_B$.  Here also the contribution to the spin
magnetic moment is mainly from the Ni d$_{3z^2-r^2}$ orbitals.

{\it Cationic CDW chains} {[Pt(chxn)$_2$][Pt(chxn)$_2$Cl$_2$]Cl$_4$}.
For PtCl the experimental 
value of the dimerization is 4.9\% (the short Pt-Cl distance 2.32
\AA~ and the long one 2.83 \AA) \cite{larsen77}. 

 The LDA + U calculation shows the absence of an
antiferromagnetic SDW solution for this system (see Table 1)
and a  dimerization of 4\%.
The 0.8 eV band gap in the dimerized state is much smaller than the
experimental gap of 1.6 eV.
This disagreement in the band gap is not surprising, since LDA + U method
works best for fairly localized orbitals with spin-orbital ordering like 
NiO \cite{ag}.  
For delocalized
orbitals, without spin polarization, like the Pt 5$d$ LDA+U reduces
essentially to the standard LDA.  We should, in principle, 
use a model where
the exchange correlation potential is non-local and dynamically
screened, for example, the GW approximation of Hedin \cite{hedin}.

{\it Anionic CDW chain} K$_4$PtI$_4$PtI$_6$.  This is a rather exotic
material among the MX-chain family in that it does not have
complicated organic ligands transversely bound to the metallic atom.
We have chosen to study K$_4$PtI$_4$PtI$_6$ (KPtI for short), because
of the simplicity of the ligands (I).  
The structure could be described as
PtI$_6$ octahedra connected along the z-axis to form a chain, with K
atoms located between the chains. While the measured structure is more
complex \cite{kpti} than the other systems computed in this Letter (including 
Pt-I-Pt angle smaller than 180$^0$ and disorder along the chain). We have
modeled KPtI as a perfect linear chain without disorder.

The computed Hubbard parameter $U$ (computed as described above)
 is 3.6 eV, and is much smaller then the
typical 7 to 10 eV values for the antiferromagnetic
late-transition-metal oxides\cite{ag}.  This small value of
$U$ allows no antiferromagnetic SDW solution to the
LDA + $U$ equations for KPtI. At computed  
 3.9\% dimerization,  The  0.37 eV energy-gap
is much smaller than the experimental value of 1.0
eV.  Like the PtCl case, the underestimation of the band gap is due to
the inadequacy of LDA + U  in  handling delocalized electrons.

Several differences between PtCl and KPtI arise from the fact that Pt-N 
distance (2.06 \AA) is much shorter than the Pt-I distance (2.65 \AA),
and Pt-Pt distance in PtCl (5.16 \AA) is much shorter than the 5.91 \AA
in KPtI. In comparing PtCl to KPtI we found that PtCl has (i) a much strong
5$d$-ligand hybridisation,  (ii) a larger Pt $d_{x^2-y^2}$-$d_{3z^2-r^2}$
energy separation,  and (iii) a larger CDW reflected by the larger
Pt$^{3-\delta} \rightarrow Pt^{3+\delta}$ charge transfer (0.29 electron
in PtCl versus 0.14 electron in KPtI). 

We have studied examples of both anionic and
cationic linear MX-chain compounds by means of a parameter-free,
mean-field, multi-orbital Hubbard model implemented around the 
 LDA.  For the cationic NiX (X= Br or Cl) chains, where
the 3$d$ orbitals are much more localized than the 5$d$ of Pt, the
predicted energy band gap is found to be very close to the
experimental one.  The model also predicts that the NiX chains have an
undimerized antiferromagnetic ground-state in agreement with
experiment.
On the other hand, in PtCl and KPtI the band gap was found to be much
smaller than the measured optical gaps.


We thank A. R. Bishop, T. J. Gammel, B. I. Swanson, G. Kanner, B.
Scott, and A. Saxena for their interest in this work.  We acknowledge
partial support provided by the Department of Energy (DOE) - Basic
Energy Sciences, Division of Materials Sciences.  Supercomputer time
was provided by the Ohio Supercomputer and by the DOE.

\newpage
\begin{figure}
\caption{ Schematic structure of a single cationic 
M(chxn)$_2$][Ni(chxn)$_2$X$_2$]X$_4$. The chxn (N$_2$C$_6$H$_{14}$
ligand structure is represented in more detail in the insert (a), and
the simplified structure used in the model calculation is represented
in the insert (b), where the cyclohexane is replaced by two H atoms.
\label{fig1}}
\end{figure}
\begin{figure}
\caption{ 
LDA total DOS and energy dispersion along the chain
direction $\Gamma Z$ in the vicinity of Fermi level (5 {\bf k}-points
are used along the $\Gamma Z$ direction).  Although only one band of
Ni $d_{3z^2-r^2}$ character crosses the Fermi level, the Peierls
instability is not effective in opening an energy band gap. This is
because the elastic anharmonic potential between Ni and Cl ions along
the chain is stronger than the electron-phonon coupling. \label{fig2}}
\end{figure}
\begin{figure}
\caption{ 
LDA+U DOS in the Ni site at the experimental 
lattice parameter decomposed into various $d$-channel symmetries: (a)
for majority spin, and (b) for minority spin 
(The spin-up of a Ni atom is the down-spin of its nearest-neighbor Ni
atoms and visa-versa).  The dark shaded area represent the Pt
$d_{3z^2-r^2}$ weight, and the light shaded area the $d_{x^2-y^2}$
weight (the occupation numbers of the $d_{3z^2-r^2}$ and $d_{x^2-y^2}$
orbitals are $ n^{\uparrow}_{d_{3z^2-r^2}} = 1.0$,
$n^{\downarrow}_{d_{3z^2-r^2}} = 0.39$, $n^{\uparrow}_{d_{x^2-y^2}} =
0.65$, and $n^{\downarrow}_{d_{x^2-y^2}} = 0.60 $).  The Hubbard term
in Eq. (2) opens the energy band gap at the Fermi level of 1.8 eV and
produces an antiferromagnetic ground state.  Almost the entire
contribution to the spin magnetic moment of 0.66 $\mu_B$ is from
d$_{3z^2-r^2}$ orbitals, whereas only 0.05 $\mu_B$ is due to the
polarization of the $d_{x^2-y^2}$ orbitals. 
\label{fig3}}
\end{figure}
\widetext
\begin{table}
\caption{
Calculated dimerization, magnetic moment, and band gap for KPtI, PtCl,
NiBr, and NiCl compared to the available experimental results indicated in
parentheses.  
}
\begin{tabular}{ccccccc}
 &M-M (\AA) & U (eV)& J (eV) & Dimerization (\%) &
 Magnetic moment ($\mu_B$)&  Band gap (eV)  \\
\noalign{\hrule}
 PtI & (5.91$^{\rm a}$) &3.6 &1.0&3.9 (4.1$^{\rm a}$)  & 0 &0.4
 (1.0$^{\rm a}$) \\
 PtCl & (5.16$^{\rm b}$) &3.5 &1.0&4.0 (4.9$^{\rm d}$)&  0 & 0.8 
(1.6$^{\rm c}$) \\
 NiBr & (5.16$^{\rm e}$) & 6.9 &0.7&0 (0$^{\rm d}$)& 0.63 & 1.6 (1.28$^{\rm
e}$) \\
 NiCl & (4.89$^{\rm e}$) & 6.8&0.7&0 (0$^{\rm e}$)  & 0.66 & 1.8 (1.9$^{\rm
e}$)\\
\end{tabular}
\label{table1}
\tablenote{~~~$^{\rm a}$Ref.~\cite{kpti}~~~$^{\rm b}$Ref.~\cite{larsen77}
~~~$^{\rm c}$Ref. ~\cite{scott92}~~~$^{\rm d}$Ref.~\cite{okamoto90}
~~~$^{\rm e}$Ref. ~\cite{nicl}}
\end{table}
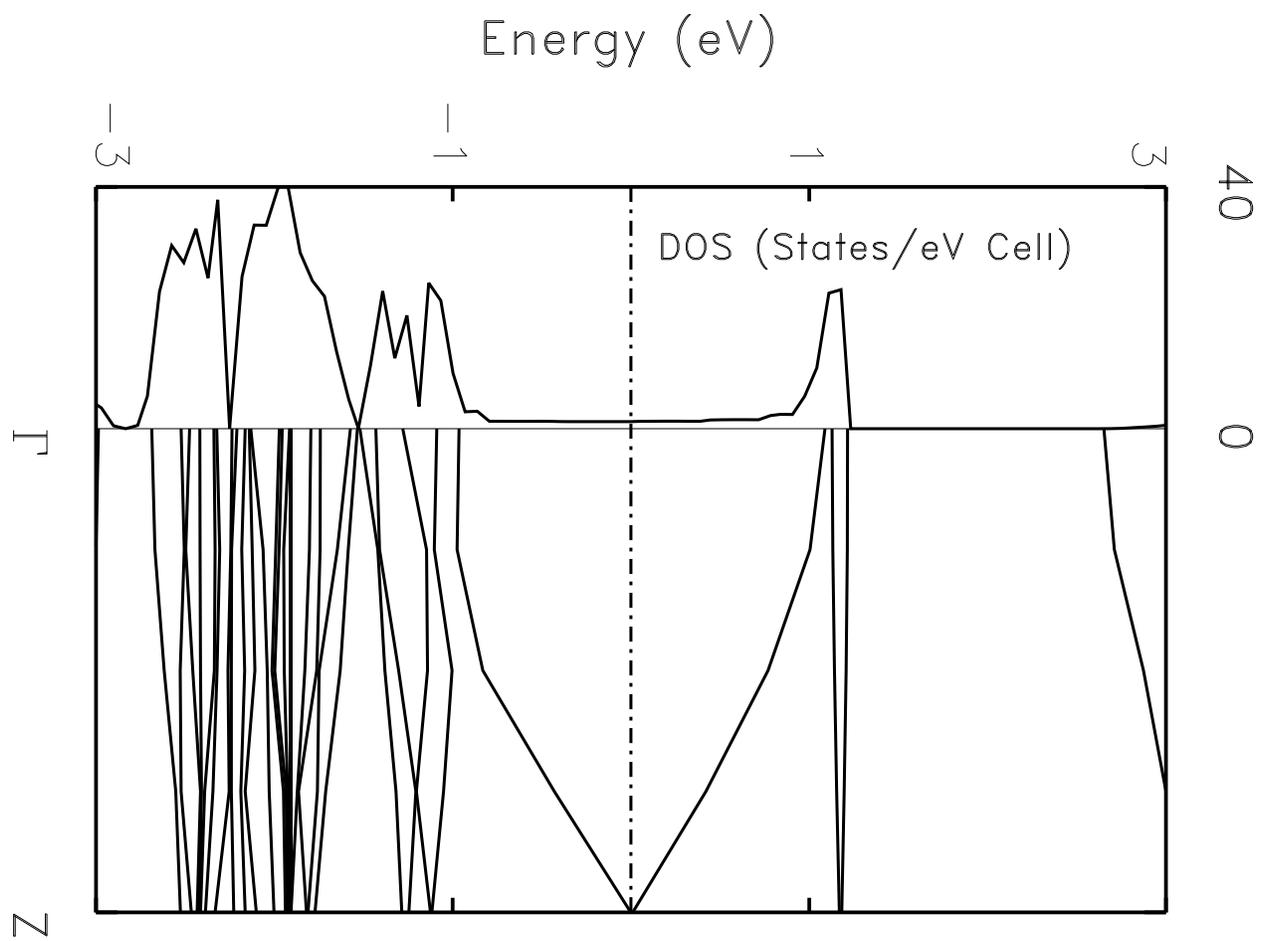
\end{document}